\documentclass[twocolumn,aps,showpacs,prl]{revtex4-1}
\usepackage{graphicx}
\usepackage{epstopdf}
\usepackage{amsmath}
\usepackage{multirow}

\begin{document}

\title{Investigation of crystal and electronic structures of MgOFeSe by first-principles calculations}

\author{Kai Liu$^{1}$}
\author{Miao Gao$^{2,1}$}
\author{Zhong-Yi Lu$^{1}$}\email{zlu@ruc.edu.cn}
\author{Tao Xiang$^{2}$}\email{txiang@iphy.ac.cn}

\affiliation{$^{1}$Department of Physics, Renmin University of
China, Beijing 100872, China}

\affiliation{$^{2}$Institute of Physics, Chinese Academy of
Sciences, P.O. Box 603, Beijing 100190, China }

\date{\today}

\begin{abstract}

In order to assist the search of new superconductors in iron selenide materials by intercalation, we calculate the crystal and electronic structures of MgOFeSe using the first-principles density functional theory. MgOFeSe is isotructural to the parent compound of iron pnictide superconductor LaOFeAs. In LaOFeAs, the anion O$^{2-}$ is located at the center of each LaO tetrahedra. But for MgOFeSe, we find that the crystal structure with the cation Mg$^{2+}$ as the tetrahedral center in the MgO layer is energetically more stable. The low energy band structures around the Fermi surfaces of MgOFeSe are contributed mainly by Fe 3$d$ orbitals. The ground state of MgOFeSe is collinearly antiferromagnetically ordered. The height of Se atoms above the Fe-Fe layer is about 1.38 \AA, which is close to the height of As from the Fe-Fe layer in the iron pnictide superconductors with optimal superconducting transition temperatures.

\end{abstract}


\maketitle


The discovery of high-$T_c$ superconductivity in LaFeAsO with partial substitution of O by F atoms \cite{Kamihara-JACS} has stimulated great interest in the study of iron-based superconductors \cite{Rotter-PRL,Wang-SSC,11}. It is commonly believed that the Fe$X$ ($X$=pnictogen and chalcogen) layers formed by the edge-shared Fe$X_4$ tetrahedras play the central role in the superconducting pairing in these materials. Among all iron-based superconductors, the PbO-type iron selenide ($\beta$-FeSe) has the simplest structure \cite{11}. But it shows very rich phase diagram under high pressure or upon alkali-metal intercalation. The superconducting transition temperature $T_c$ of FeSe is about 8 K at ambient pressure \cite{11}, and can be raised up to 37 K by applying high pressure \cite{gpa1,gpa2,gpa3}.
The Potassium or other alkali-metal intercalated FeSe becomes superconducting above 30 K \cite{chenxl,Maziopa-JPCM,Fang-EPL,Yan-PRL}, or even at 44 K with excess Fe \cite{zhang13} or at 48 K under high pressure \cite{SunLiLing}.
Signatures of  superconductivity above 50 K have also been observed in the FeSe monolayer grown on the substrate SrTiO$_3$ by photoemission and transport measurements \cite{xue,zhou1,zhou2,feng}.

Intercalation to FeSe, or more generally to other layered materials, provides an experimentally feasible route to find new superconductors. A material which has recently attracted interest is the MgO intercated FeSe\cite{chen2013}. In this material, a layer MgO is used to be a spacer intercalated between two FeSe layers. Unlike in the alkali-metal intercalated FeSe, MgO is charge-neutral and does not serve as a charge reservoir for the FeSe layers.

\begin{figure}[h]
\begin{center}
\includegraphics[width=8.0cm]{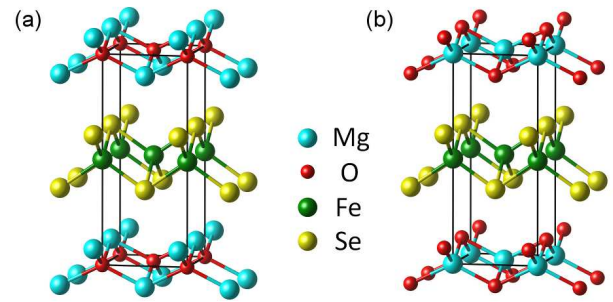}
\caption{(Color online) Two kinds of crystal structures of MgOFeSe with  P4/nmm space group symmetry. The difference between these two structures lies whether the MgO layer is composed of edge-sharing tetrahedra with an O center (panel a) or an Mg center (panel b). The thick black line denotes the unit cell.
}
\label{figa}
\end{center}
\end{figure}

In this work, we present a first principles density functional theory study on the atomic and electronic structures of MgOFeSE.
MgOFeSe is isostructural to LaOFeAs, whose space group is P4/nmm (No. 129). Two different crystal structures of MgOFeSe, as shown in Fig.~\ref{figa}, are considered. The difference between these two structures lies whether the MgO layer is composed of edge-sharing tetrahedral with an O center [Fig.~\ref{figa}(a)] or an Mg center [Fig.~\ref{figa}(b)].
We find that structure (b) has a lower energy than structure (a) in both the non-magnetic and magnetic ordered phases. This indicates that structure (b) is energetically more stable than structure (a), different from the case in LaOFeAs where the center of LaO tetrahedra is occupied by anionic oxygen. Similar as in undoped LaOFeAs, we find that the non-superconducting ground state of MgOFeSe is collinearly antiferromagnetically (AFM) ordered.

The first-principles electronic structure calculations were implemented with the VASP package \cite{ab1,ab2}, which makes use of the projector augmented wave (PAW) method \cite{paw}.
The exchange-correlation functional was represented by the generalized gradient approximation (GGA) given by Perdew-Burke-Ernzerhof (PBE) \cite{pbe}. To describe the interlayer van der Waals (vdW) interactions not included in the conventional density functional, both the semi-empirical DFT-D2 method \cite{Grimme} and a more accurate vdW-optB86b functional \cite{Klim,Klim2} were adopted. The energy cutoff for the plane waves was set to 520 eV, which is high enough to ensure the accuracy for variable cell relaxations. The $1 \times 1 \times 1$ tetragonal cell of MgOFeSe was used for the nonmagnetic and the AFM N\'eel states and the integrations over the Brillouin zone were performed with a 12 $\times$ 12 $\times$ 12 {\bf k}-point mesh. The $\sqrt{2} \times \sqrt{2} \times 1$ tetragonal cell and an 8 $\times$ 8 $\times$ 8 {\bf k}-point mesh were used for the collinear AFM state. We found that the {\bf k}-points along the $z$ direction are dense enough to describe the interlayer interactions. A gaussian smearing method with a width of 0.05 eV was used for the Fermi-level broadening. Both the cell parameters and the internal atomic positions were allowed to relax.
The convergence criterion for the forces on atoms was smaller than 0.01 eV/\AA.

\begin{table}
\caption{Calculated fully-relaxed lattice parameters for MgOFeSe in the AFM N\'eel state with and without vdW interactions. As a comparison, the measured parameters for bulk FeSe at 298 K are also given.}
\begin{center}
\begin{tabular*}{8.3cm}{@{\extracolsep{\fill}} cccccccc}
\hline
\hline
$ $ & $a=b$(\AA) & $c$(\AA) & $h_{Se}$(\AA) \\
\hline
PBE & 3.8430 & 9.9056 & 1.393 \\
DFT-D2 & 3.8094 & 8.5644 & 1.383 \\
vdW-optB86b & 3.8002 & 8.9122 & 1.381 \\
\hline
FeSe (298 K)\cite{mcqueen} & 3.7734 & 5.5258 & 1.476 \\
\hline \hline
\end{tabular*}
\end{center}
\end{table}

\begin{table}[!b]
\caption{Calculated fully-relaxed lattice parameters and relative energies $E_r$ in three different phases with respect to the energy of the non-magnetic phase for MgOFeSe obtained with the vdW-opt86b functional. }
\begin{center}
\begin{tabular*}{8.3cm}{@{\extracolsep{\fill}} cccccccc}
\hline
\hline
$ $ & $a$(\AA) & $b$(\AA) & $c$(\AA) & $h_{Se}$(\AA) & $E_r$(meV/Fe) \\
\hline
Nonmag. & 3.7844 & 3.7844 & 8.8229 & 1.322 & 0 \\
N\'eel & 3.8002 & 3.8002 & 8.9122 & 1.381 & -27.9 \\
Collinear & 3.8264 & 3.8129 & 8.8681 & 1.383 & -55.1 \\
\hline \hline
\end{tabular*}
\end{center}
\end{table}

We calculated the electronic structures for the two kinds of crystal structures of MgOFeSe shown in Fig.~\ref{figa}. The total energy of structure (b) is found to be about 90-120 meV/Mg lower than that of structure (a) in all the phases we have examined, including the non-magnetic, the AFM N\'eel, and the collinear AFM phases (schematically shown in Fig. 2).
This indicates that structure (b) is energetically more stable for MgOFeSe, as stated before.
At a glance one may think that structure (a) should have a lower energy because in structure (a) the cation atom Mg in the MgO layer lies closer to the anion atom Se in the adjacent FeSe layer, which would generally reduce the potential energy. This intuitive argument is indeed valid for LaOFeAs and other iron pnictide materials since the FeAs layer is negatively charged, but not valid for MgOFeSe since both the FeSe and MgO layers are charge-neutral. In MgOFeSe, the charge-neutral nature of MgO and FeSe layers ensures that there is no charge transfer between these layers. This suggests that the energy difference between the two structures should come mainly from the MgO layer and the vdW interactions may have significant contribution to the interlayer bonding. In order to understand why structure (b) is more stable than structure (a), we calculated the total energy for a freestanding MgO single layer with an in-plane lattice constant of 3.7872 \AA. We found that the energy of the isolated MgO single layer in structure (b) is lower than that in structure (a) by 208 meV/Mg. This ascertains that the energy difference between structures (a) and (b) for MgOFeSe indeed originates from the MgO spacer layer. Thus we adopt structure (b) as the crystal structure of MgOFeSe in the following discussion.

\begin{figure}
\includegraphics[width=8.2cm]{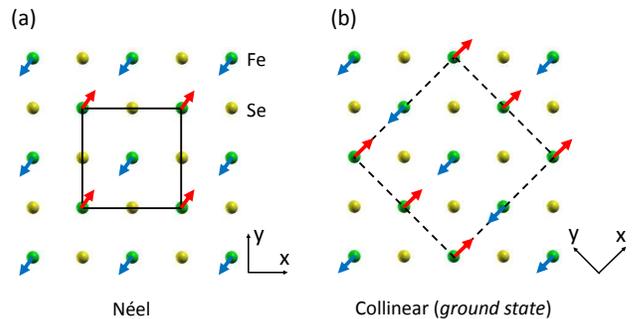}
\caption{(Color online)
Schematic top views of two magnetic orders in the Fe-Fe layer: (a) the conventional checkerboard antiferromagnetic N\'eel order and (b) the collinear antiferromagnetic order (which is the ground state of MgOFeSe). The solid (a) and dashed (b) squares denote the magnetic unit cells. }\label{figb}
\end{figure}

Table I reports the lattice parameters of MgOFeSe in the AFM N\'eel state calculated with (DFT-D2 and vdW-optB86b respectively) and without considering the vdW corrections. We find that a good agreement between the DFT-D2 and the vdW-optB86b calculations. We have also evaluated the lattice parameters in the non-magnetic and other AFM phases using the vdW-optB86b functional. The results are given in Table II. According to the relative energies of MgOFeSe in different magnetic orders (Table II), the ground state of MgOFeSe is found to have the collinearly antiferromagnetic order, same as for bulk FeSe \cite{mafengjie}.


\begin{figure}
\includegraphics[width=8.6cm]{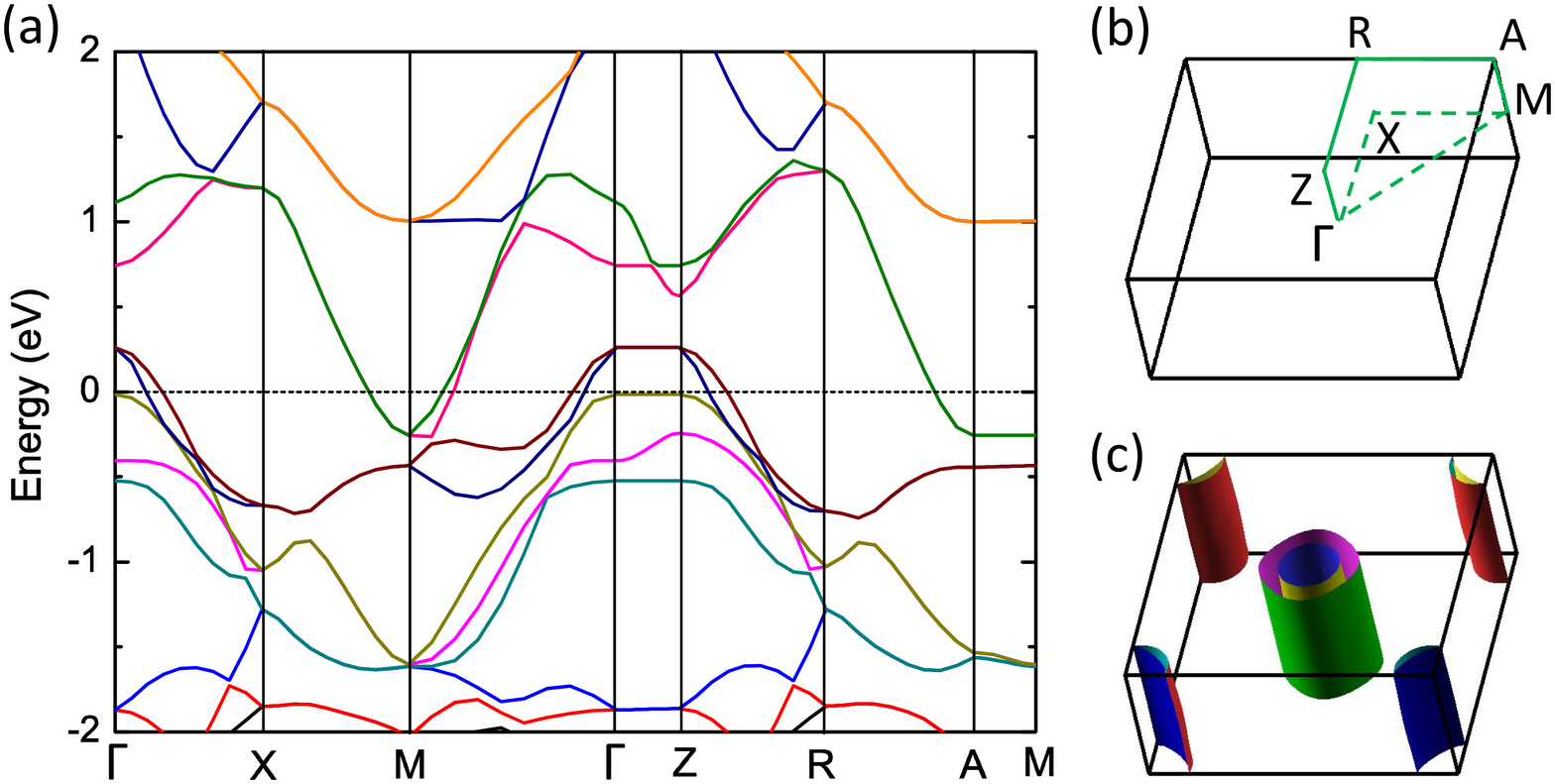}
\caption{(Color online) Electronic structure of MgOFeSe with the structure (b) [Fig.~\ref{figa}(b)] in the non-magnetic state: (a) the band structure, (b) the Brillouin zone, and (c) the Fermi surface. The Fermi energy is set to zero. }\label{figc}
\end{figure}

\begin{figure}[!b]
\includegraphics[width=8.0cm]{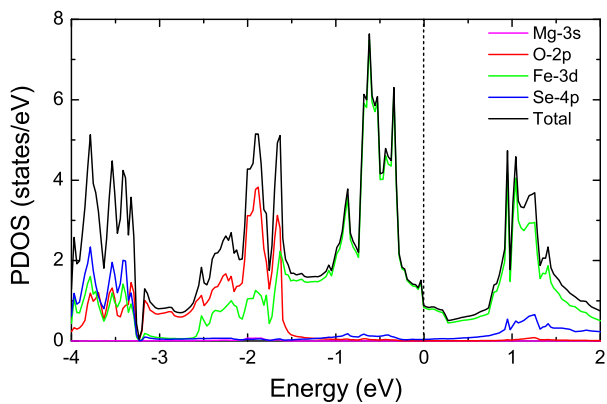}
\caption{(Color online) Calculated atomic orbital-resolved partial
density of states per formula of MgOFeSe in non-magnetic state. The Fermi energy is set to zero.}
\label{figd}
\end{figure}

Figure~\ref{figc} shows the electronic band structure and the Fermi surface for MgOFeSe in the non-magnetic state. The electronic structures in the non-magnetic state provide a reference for the study of magnetic states. In the non-magnetic phase, there are two bands crossing the Fermi level, forming two hole-like cylinders of Fermi surfaces along $\Gamma$-Z line and two electron-like cylinders of Fermi surfaces around the zone corners [Fig.~\ref{figc}(c)]. This is similar to the case for the bulk FeSe \cite{Subedi, Tamai}. The band dispersion along the $\Gamma-Z$ and $M-A$ directions are very weak,  demonstrating the two-dimensional characteristics. The calculated density of states (DOS) is shown in Fig.~\ref{figd}. Basically the DOS separates into two parts. Below -1.5 eV, the bonding states between O and Mg atomic orbitals have the largest contribution to the DOS. Between -1.5 eV and 1.5 eV, the DOS consists mainly the contribution from the Fe-$3d$ orbitals. Thus the low energy physics is dominated by the Fe $3d$ electrons.

\begin{figure}
\includegraphics[width=8.6cm]{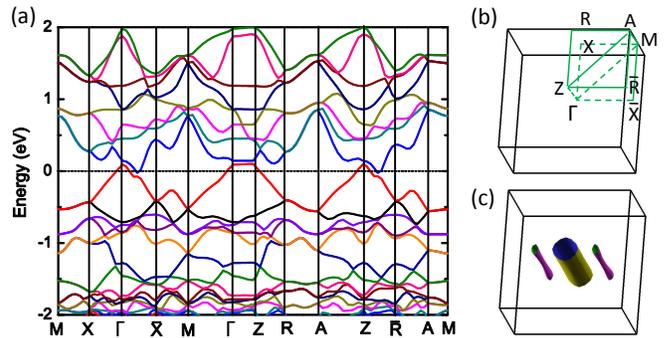}
\caption{(Color online) Electronic structure of MgOFeSe with the structure (b) [Fig.~\ref{figa}(b)] in the collinear antiferromagnetic state: (a) the band structure, (b) the Brillouin zone, and (c) the Fermi surface. The Fermi energy is set to zero. $\Gamma$-$X$ and $\Gamma$-$\bar{X}$ correspond to the parallel-aligned and antiparallel-aligned magnetic moment lines, respectively.
} \label{fige}
\end{figure}

The collinearly AFM-ordered state, as shown in Fig. \ref{figb}(b), is a common ground state of the parent compounds of iron-based superconductors \cite{Cruz-Nature,Huang-PRL}, resulting from the anion-bridged AFM superexchange \cite{Ma-PRB}. The electronic band structures of MgOFeSe in the collinear AFM state are shown in Fig.~\ref{fige}. In this phase, similar as in LaOFeAs\cite{Ma-PRB}, there is a strong depletion of the energy bands around the Fermi level, and only one hole pocket around the $\Gamma$ point and two electron pockets along the $\Gamma-\bar{X}$ direction survive. The vdW functional has strong effect on the lattice constant along the stacking direction, but it does not affect much on the electronic structures.

\begin{figure}
\includegraphics[width=8.3cm]{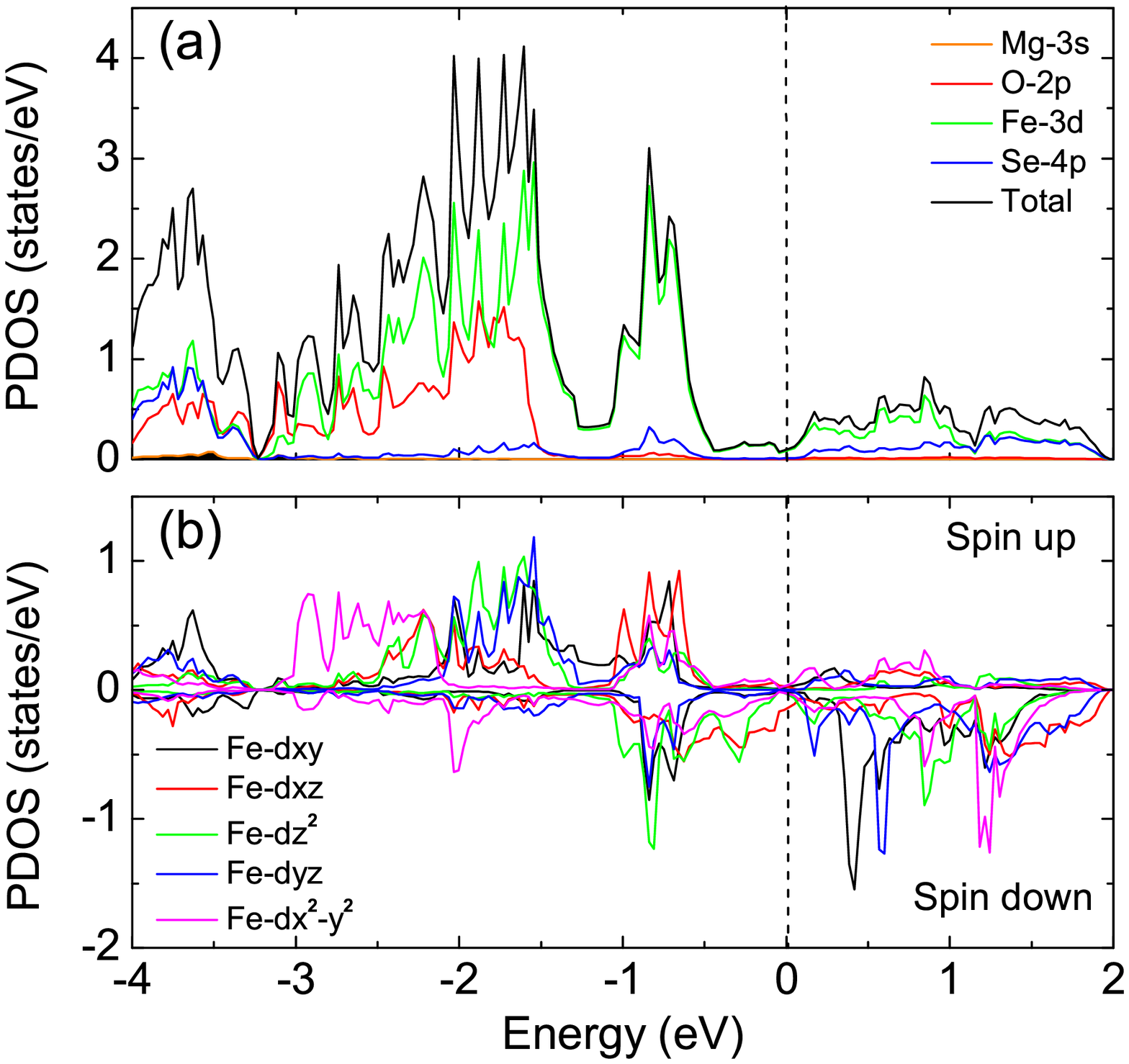}
\caption{(Color online) (a)
Total and orbital-resolved partial densities of states per formula (spin-up part) of MgOFeSe
in the collinear AFM state. (b) Projected densities of states for the five $3d$ orbitals around one up-spin Fe atom. The Fermi energy is set to zero.}
\label{figf}
\end{figure}

Figure \ref{figf} shows the calculated total and projected densities of states
in the collinear AFM state.
In comparison with the nearly completely filled five up-spin orbitals, the five down-spin orbitals are only partially filled. However, the down spin electrons are nearly uniform distributed among these five $3d$ orbitals, which indicates that there is strong hybridization between them and the crystal-field splitting imposed by the Se atoms is small.

\begin{figure}
\includegraphics[width=8.6cm]{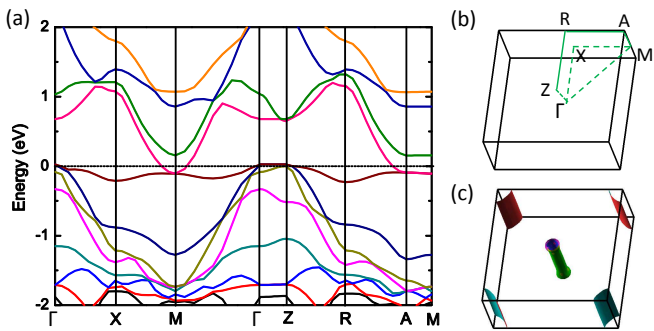}
\caption{(Color online) Electronic structure of MgOFeSe with the structure (b) [Fig.~\ref{figa}(b)] in the checkerboard antiferromagnetic N\'eel state: (a) the band structure, (b) the Brillouin zone, and (c) the Fermi surface. The Fermi energy is set to zero.} \label{figg}
\end{figure}

Experimentally bulk FeSe at ambient pressure is found to be in a paramagnetic state due to extra Fe atoms \cite{gpa,nmr,bendele}. Considering that the AFM checkerboard N\'eel order and the paramagnetic phase share many important features (such as local moments around Fe atoms and zero net magnetic moments in a unit cell), the AFM N\'eel state can be adopted to model the paramagnetic phase in many aspects \cite{Bazhirov, JiWei}. Fig.~\ref{figg}  shows the band structure and the Fermi surface of MgOFeSe in the AFM N\'eel state. In this state, the hole pockets around the $\Gamma$ point are very small and they can be easily eliminated by electron doping. The dispersionless energy band right below the Fermi level is consistent with previous calculations for both the bulk FeSe and the isolated FeSe single layer in the AFM N\'eel order \cite{Bazhirov,Bazhirov2}.
For the FeSe monolayer grown on SrTiO$_3$\cite{xue}, only the electron pockets around the $M$ point were observed in the angle-resolved photoemission spectroscopy (ARPES) measurements \cite{zhou1,zhou2,feng}. This Fermi surface feature can be reproduced by the density functional theory only in the AFM N¡äeel state\cite{Bazhirov2,zhangping, Liu-PRB}.

Empirically, it is found that the superconducting transition temperature $T_c$ of iron-based superconductors has strong correlation with the height of anion ($h_{\text{Se}}$ or $h_{\text{As}}$) above Fe-Fe square lattice. The highest $T_c$ often occurs when $h_{\text{Se}}$ or $h_{\text{As}}$ is about 1.38 {\AA} \cite{gpa3,Mizuguchi-SST}. Our calculated $h_{\text{Se}}$ of structure (b) in the magnetic states is about 1.38 {\AA}, close to the optimal value at which the highest-T$_c$ is found in iron pnictide
superconductors.

In summary, from the energetic point of view, our first-principles calculations with van der Waals corrections on MgOFeSe suggest that a crystal structure with Mg as the tetrahedral center in MgO layer is energetically favorable. The ground state is collinearly antiferromagnetic ordered. The electronic band structure near the Fermi level and the Fermi surfaces of MgOFeSe are determined by the FeSe layer. The height of Se from the Fe-Fe plane is found to be $\sim$1.38 \AA, close to the height of As from the Fe-Fe layer in the iron pnictide superconductors with optimal T$_c$, which suggests that MgOFeSe is a good candidate of iron-based high-T$_c$ superconductor.
The effective role played by the MgO spacer layer in tuning the delicate structure of FeSe layers can be also found in other intercalated FeSe compounds.


This work is supported by National Natural Science Foundation of China (Grant Nos. 11004243, 11190024, and 51271197) and National Program for Basic Research of MOST of China (Grant No. 2011CBA00112). Computational resources have been provided by the Physical Laboratory of High Performance Computing at Renmin University of China. The atomic structures and Fermi surfaces were prepared with the XCRYSDEN program \cite{kokalj}.

\end{document}